\documentclass[aps,prd,preprintnumbers,nofootinbib,twocolumn]{revtex4}
\usepackage{bm}
\usepackage{latexsym}
\usepackage{dcolumn}
\usepackage{amsmath,amsfonts,amssymb}
\usepackage{graphicx,epsfig}
\usepackage{psfrag}
\usepackage{amsthm}

\def\be {\begin{equation}}
\def\ee {\end{equation}}
\def\bea {\begin{eqnarray}}
\def\eea {\end{eqnarray}}
\def\bc {\begin{center}}
\def\ec {\end{center}}
\def\bfg {\begin{figure}}
\def\efg {\end{figure}}
\def\bi {\begin{itemize}}
\def\ei {\end{itemize}}

\def\no {\noindent}

\def\b  {\beta}

\def\d  {\delta}

\def\beq{\begin{equation}}
\def\eeq{\end{equation}}
\def\br{\begin{eqnarray}}
\def\er{\end{eqnarray}}
\newcommand{\eel}[1] {\label{#1}\end{equation}}

\newcommand{\bdm}{\begin{displaymath}}
\newcommand{\edm}{\end{displaymath}}

\begin{document}
\title{Reply to ``Comment on `Universality of Quantum Gravity Corrections'~"}


%

\author{Saurya Das}
\email{saurya.das@uleth.ca}

\affiliation{Department of Physics, University of Lethbridge,\\
 4401 University Drive, Lethbridge, Alberta T1K 3M4, Canada}

\author{Elias C. Vagenas}
\email{evagenas@academyofathens.gr}

\affiliation{Research Center for Astronomy and Applied Mathematics,\\
Academy of Athens, \\
Soranou Efessiou 4,
GR-11527, Athens, Greece}

\begin{abstract}
We address the three points raised by the authors of the above Comment (ref.\cite{ette}). 
\end{abstract}

\maketitle



\begin{enumerate}

\item{Point 1}\\
In our paper \cite{Das:2008kaa}, we showed that depending on the value of the 
GUP parameter $\beta$, a typical STM could register an excess of one electron charge ($1e$) 
due to quantum gravity effects, in about a year. 
The authors of \cite{ette} claim that to actually be able to measure this effect, 
one would need to have a circuit with frequency 
$f 
\sim 10^{19}~Hz$.
Their estimate appears to
arise from the misunderstanding that one would need the accuracy to measure one electron charge 
in a current of $1A$ (i.e. in $1$ second),
corresponding to an accuracy of $1$ part in $10^{19}$. We point out that this is 
not the case. One would simply need the apparatus to measure electric charge with an 
accuracy of $1e$, {\it not in $1$ second}, but 
in any reasonable amount of time, which can surely be done (and indeed has
been possible since the time of Millikan) \cite{millikan}.

\item{Point 2}\\
Varying the standard expression for the STM current $I$ 
(proportional to the transmission coefficient $T$)
with respect to the gap $a$ between the needle and sample
(measurable currently to an accuracy of about $10^{-15}m$),
and together with Eq.(32) of our paper \cite{Das:2008kaa},
the authors of \cite{ette} claim that in effect $\d I/I \approx 10^{-10}$. 
There seems to be at least two errors in this interpretation: \\
(i) one should vary the {\it GUP corrected current}, proportional to Eq.(30) of ref.\cite{Das:2008kaa}, which gives
$\frac{\Delta I}{I} \sim - k_1 \Delta a + \beta_0 \ell_{pl}^{2} k_{1}^{3}\Delta a$.
Clearly, the last term being much smaller is the relevant one, and when this is 
combined with Eq.(32) of \cite{Das:2008kaa}, $\b_0$ cancels from both sides, and 
no bound on the latter is obtained. \\
(ii) Surfaces are imaged in an STM in two ways,
the {\it constant height mode}, in which $a$ and the voltage $V$ are held fixed, while
$I$ changes, and the {\it constant current mode}, in which the $I$ is held fixed and 
$a$ varies. The former being a faster method is often preferred, and our calculations
{\it per se} pertain to this mode, in which the variation of $a$, and
its available accuracy of measurement are irrelevant.

\item{Point 3}\\
The authors claim that statistical errors could be important in STM measurements. 
This may indeed be the case. However, 
we would like to remind the readers that our analysis was intended to show
that in principle the GUP can affect well understood quantum mechanical
systems such as the STM. If actual experiments to measure
these effects are planned, one would of course have to take into
account many such sources of error and other tiny physical effects as
well. Furthermore, their bound of $\beta_0 > 10^{39}$
is based on their assumed accuracy of measurement of the current, time etc,
whereas much better measurements already exist. Thus this bound
does not seem to be robust.

\end{enumerate}

\no {\bf Acknowledgment}

This work was supported in part by the Natural
Sciences and Engineering Research Council of Canada and by the
Perimeter Institute for Theoretical Physics.



\end{document}